\documentclass[12pt,preprint]{aastex}

\newcommand\mes{M{\'e}sz{\'a}ros}

\begin{document}

\shorttitle{Early X-ray afterglows}
\shortauthors{Lazzati, Begelman \& Campana}
\title{Thick fireballs and the steep decay in the early X-ray
afterglow of gamma-ray bursts}

\author{Davide Lazzati and Mitchell C. Begelman\altaffilmark{1}}
\affil{JILA, 440 UCB, University of Colorado, Boulder, CO 80309-0440,
  USA} 
\email{lazzati;mitchb@colorado.edu} 
\altaffiltext{1}{Department of Astrophysical and Planetary Sciences,
University of Colorado at Boulder}

%

\begin{abstract}
We study the early afterglows of gamma-ray bursts produced by
geometrically thick fireballs, following the development of the
external shock as energy is continually supplied to the shocked
material. We study the dependence of the early afterglow slope on the
luminosity history of the central engine. The resulting light curves
are modeled with power-law functions and the importance of a correct
choice of the reference time $t_0$ is investigated. We find that
deviations from a simple power-law are observed only if a large
majority of the energy is released at late times. The light curve in
this case can be described as a simple power-law if the reference time
is set to be close to the end of the burst. We applied our analysis to
the cases of GRB~050219a and GRB~050315. We show that the early steep
decay of the afterglow cannot result from the interaction of
the fireball with the ambient medium. We conclude that the early X-ray
afterglow emission is associated with the prompt phase and we derive
limits on the radius at which the prompt radiation is produced.
\end{abstract}
\keywords{gamma-rays: bursts --- hydrodynamics --- radiation
processes: non-thermal}

\section{Introduction}

The launch of the Swift mission (Gehrels et al. 2004) has opened a new
window in the observation of gamma-ray bursts (GRBs) and their
afterglows. The X-ray telescope (XRT) on board Swift has provided, and
is providing, high quality observations of the early afterglow
phase. In particular it has been possible, for the first time, to
follow the transition between the prompt and the afterglow emission at
high energy (Tagliaferri et al. 2005; Chincarini et al. 2005). X-ray
observations are important for studying this phase of the burst since the
high energy emission is less heavily contaminated by reverse shock
emission (Akerlof et al. 1999; Sari \& Piran 1999).

Swift observations have revealed that in more than half the events the
X-ray emission at early times is characterized by a very steep decay
(Chincarini et al. 2005; Tagliaferri et al. 2005). This initial phase
is followed by a slower decay. At later times, the afterglow light
curve steepens again, possibly in association with the transition from
a spherical to a beamed behavior (see Panaitescu et al. 2005 and Zhang
et al. 2005 for a thorough discussion of the late time breaks). Broken
power-law (BPL) fits to the early light curves yield early decay
indices $-4\lesssim\delta_1^X\lesssim-2.5$. It has been noted that the
slopes depend strongly on the time $t_0$ that is used as a reference
from which all times are measured (Piro et al. 2005; Tagliaferri et
al. 2005; Panaitescu et al. 2005; Zhang et al. 2005). The power-law
behavior of the light curve is due to the self-similar evolution of
the radius of the external shock with time. This is an accurate
approximation if the time spent by the reverse shock in crossing the
shell is short compared to the dynamical time (e.g., Sari 1997). If the
reverse shock takes time to cross the fireball, the energy injection
into the external shock is not impulsive and a self-similar solution
(i.e., power-law behavior) is attained only at later stages. For this
reason, equating the beginning of the $\gamma$-ray emission
($t_\gamma$) to $t_0$ can lead to an overestimate of the decay slope
of the early afterglow.

In this paper we compute external shock light curves from
geometrically thick fireballs with a variety of luminosity
histories. We use these simulated light curves to explore the
dependence of $t_0$ on the energy ejection history of the central
engine that powers the fireball. We show that, in most cases, assuming
$t_0=t_\gamma$ is a good approximation and that $t_0>t_\gamma$ should
be considered only if the central engine luminosity is a steeply
increasing function of time (\S~2).  In \S~3 we use the light curves
of GRB~050219A and GRB050315 as test cases. We show that their early
light curves can be described either as the sum of an exponential
decay plus a power-law (EXP+PL) or as the sum of two power-laws
(PL+PL). We argue that the steepness of the decay cannot be accounted
for by any process related to the external shock (for which
$t_0=t_\gamma$). We conclude that the most likely origin of the
observed steep decays is large angle radiation from the prompt phase
(Kumar \& Panaitescu 2000; Zhang et al. 2005), and we derive stringent
lower limits on the radius at which the prompt emission is released.

\section{The $t_0$ of thick shells}

Suppose the GRB central engine releases a luminosity $L=L(t)$ between
the times $t_\gamma$ and $t_\gamma+T_{\rm{GRB}}$. Since the motion of
the fireball into the external material is extremely supersonic, a
shock structure develops with a forward shock propagating outward into
the interstellar medium (ISM) and a reverse shock propagating backward
into the ejecta. At late times, a self-similar solution is obtained,
where the shock properties (Lorentz factor, radius, temperature,
density) scale as power-laws of the observed time (Blandford \& McKee
1976). Such a solution is valid only if the time scale during which
the energy is supplied to the shock region is
negligible\footnote{Different self-similar solutions are obtained if
the energy is supplied as a power-law in time.}. Such an approximation
is usually excellent for the late afterglow emission (unless some
delayed activity of the central engine is involved, e.g., Panaitescu
et al. 1998).

Here we consider the very early afterglow of long GRBs, during the
time interval in which the energization of the external shock by the
ejecta is taking place. It has been shown and observed that this
energization can be accompanied by bright optical emission: the
optical flash produced by the reverse shock crossing the ejecta
(Akerlof et al. 1999; Nakar \& Piran 2004). To study how the finite
duration of the energy injection phase affects the afterglow, we
numerically compute light curves from the external shock. The interval
($T_{\rm{GRB}}$) during which the central engine is active is divided
into sub-intervals. The first part of the ejecta is allowed to
interact with the ISM and to decelerate according to energy and
momentum conservation. The subsequent parts freely propagate in a
vacuum until they catch up with the shocked material. At this point
the energy and momentum of the shock is increased by the added amounts
and the dynamics is modified accordingly. Since we are interested in
the dynamics of the process and its effect on self-similarity we do
not compute the emission from the reverse shock but only the emission
from the forward shock. The treatment presented here does not consider
short-time transient phenomena related to the propagation of the newly
injected energy from the contact discontinuity to the forward
shock. Such effects could be important in the case of discontinuous
energy ejection. In that case, a complex structure of reflected and
propagated shocks could form. This structure should form a variability
pattern on top of the envelope of the light curve we compute here. We
plan to explore these transient phenomena in a future work based on
hydrodynamical simulations.

\begin{figure}
\plotone{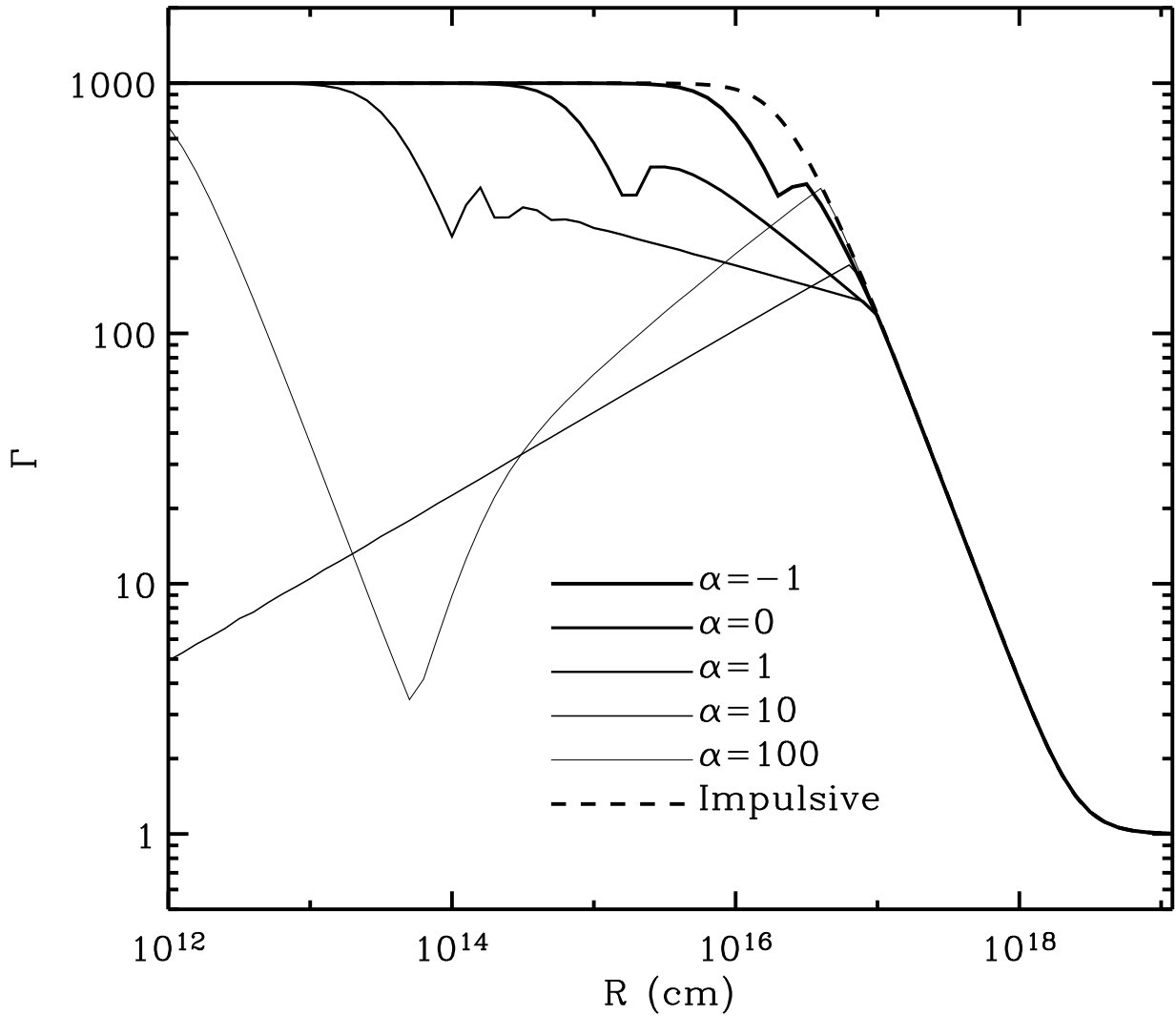} 
\figcaption{Evolution of the Lorentz factor of the shocked material as
a function of radius for 6 different energy ejection histories,
parametrized through the index $\alpha$ of the power-law
$L(t)\propto{t}^\alpha$. For the $\alpha=-1$ lightcurve, $L(t)$ is
constant between $t=0$ and $t=1$~s in order to avoid divergence. The
dashed line shows the impulsive approximation, while the solid lines
show different power-law behaviors. See Fig.~\ref{fig:lc} for the
relative GRB light curves and afterglows. Some residual numerical
noise is present at the deceleration radius of the first subsection of
the fireball. We checked by varying the resolution that this numerical
noise does not affect the resulting light curves.
\label{fig:gr}}
\end{figure}

Figure~\ref{fig:gr} shows the evolution of the Lorentz factor of the
shocked material as a function of the radius of the external shock for
various energy ejection histories of the central engine.  We
parametrize these light curves as power-laws in time,
$L(t)\propto{}t^{\alpha}$.  A dashed line shows the evolution of the
Lorentz factor in the impulsive energization approximation, while
lines of different thickness show the 5 different luminosity
histories. As expected, the asymptotic behaviors are similar, but a
transient phase that deviates from the self-similar solution is
observed at early and intermediate times. After the dynamics has been
calculated, we compute light curves for the external shock using the
code developed by Rossi et al. (2004; see also Rossi et al. 2002).

\begin{figure}
\plotone{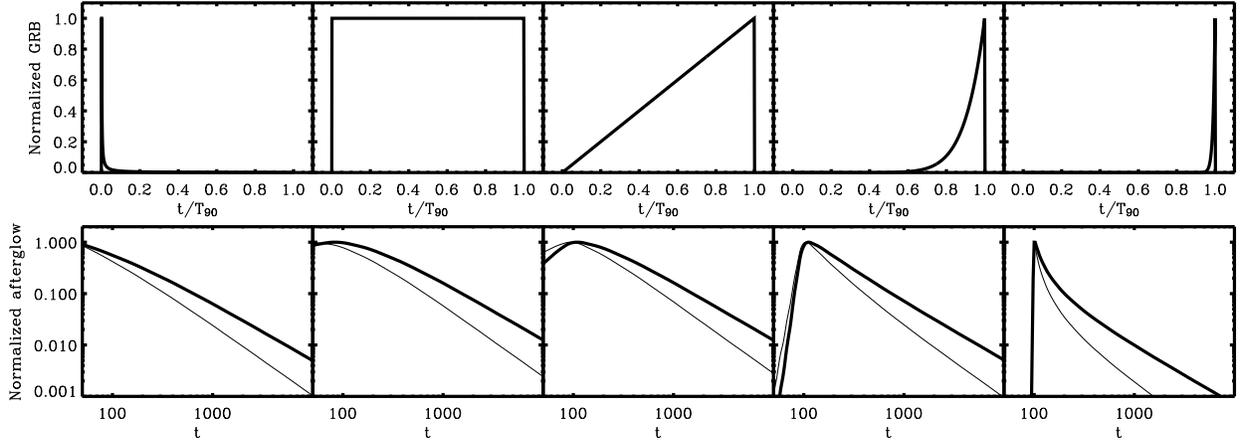} 
\figcaption{Light curves of the prompt (upper panels) and afterglow
emission (lower panels) from GRBs characterized by thick fireball
shells. We consider light curves of the form
$L\propto{t}^\alpha$. From left to right, upper panels show the light
curve of the central engine (for a constant efficiency of conversion
into radiation) for $\alpha=-1$, 0, 1, 10, 100. The lower panels show
the corresponding optical light curves (thick lines) and X-ray light
curves (thin lines). For this particular simulations
$T_{\rm{GRB}}=100$~s has been assumed. All light curves are normalized
to their maximum value. The origin of times has been set as
$t_0=t_\gamma$ in all cases. Note that only for $\alpha\ge10$ does
this choice lead to an overestimate of the early slope of the
afterglow. A constant entropy jet is assumed in all cases. See Rossi
et al. (2004) for details on the afterglow computation.
\label{fig:lc}}
\end{figure}

In Figure~\ref{fig:lc} we show the results of our computations for a
set of luminosity curves of the central engine.  We find that the self
similar behavior of the external shock emission is well-described by a
power-law if $t_0=t_\gamma$ in all but the most extreme cases.  Only
the GRB with light curves indices $\alpha\ge10$ require $t_0>t_\gamma$
in order to model the decaying part of their light curves as simple
power laws. Note that in these computations the spectrum of the
afterglow emission is simplified as a single power-law with
$F(\nu)\propto\nu^{-(p_e-1)/2}$ for optical light curves and
$F(\nu)\propto\nu^{-p_e/2}$ for X-ray light curves. Here $p_e$ is the
index of the electron distribution and has been numerically set to
$p_e=2.5$.

To constrain formally the value of $t_0$ we fit the decaying part of
the light curves with a function of the form:
\begin{equation}
F(t) = A \left(t-\frac{t_0-t_\gamma}{T_{\rm{GRB}}}\right)^{-\delta}.
\end{equation}
In most cases, due to the smooth roll-over of the increasing part of
the afterglow to the decaying part, a formal $t_0<t_\gamma$ is
obtained. Formal fit results are detailed in Table~\ref{tab:1}.

The presented results do not depend on the choice of the value of
$T_{\rm{GRB}}$ as long as the thick shell condition is preserved. This
condition was derived by Sari \& Piran (1995) and reads\footnote{Note
that this condition is strictly valid for a constant luminosity of the
central engine ($\alpha=0$).}:
\begin{equation}
\xi\equiv\left(\frac{E}{3\pi\,n\,m_p\,c^5\,T_{\rm{GRB}}^3\Gamma_0^8}
\right)^{1/6} < 1,
\end{equation}
where $n$ is the density of the external material, and $\Gamma_0$ the
asymptotic Lorentz factor of the fireball.  We have also checked that
the result does not depend on the level of refinement in the
discretization of the GRB light curve.

\section{Case studies: GRB~050219A and GRB~050315}

\begin{figure}
\plotone{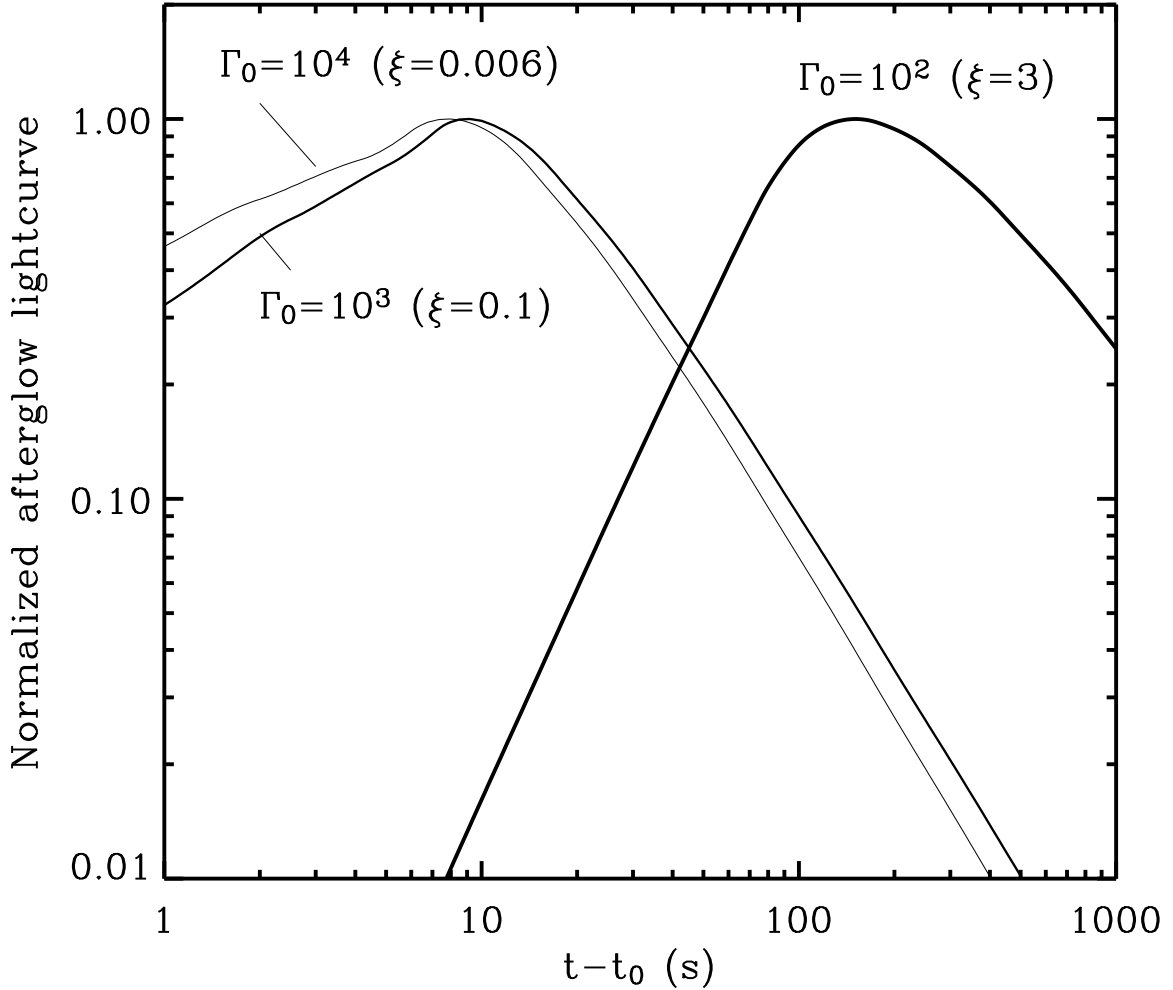}
\figcaption{X-ray afterglow light curves for GRB~050219A with a
continuous energy injection that tracks the prompt emission light
curve, observationally determined assuming $z=1$. Three different
initial Lorentz factor of the fireball are considered: $\Gamma=100$,
1000 and $10^4$. In the first case the deceleration time is longer
than the energy injection time and there is no modification with
respect to the impulsive injection approximation. In the higher
Lorentz factor case, the shell is thick and the light curves deviate
from the impulsive approximation.
\label{fig:lc0219}}
\end{figure}

We first consider GRB~050219A as a test case, due to the quality of
its early X-ray light curve (Tagliaferri et al. 2005). Due to the lack
of redshift measurements for this event, we assume a redshift
$z=1$. This choice is made for clarity and is used in the computation
of the light curves but does not affect any of the dynamical
considerations. We assume that the efficiency of the conversion of
kinetic energy to radiation is constant throughout the GRB, so that
the light curve tracks the luminosity of the central engine (see Fig,
2 of Tagliaferri et al. 2005). We compute the light curves of the
early forward shock emission as described in
\S~2. Figure~\ref{fig:lc0219} shows the resulting light curve for
different choices of the fireball dimensionless entropy
($\eta=L/(\dot{M}\,c^2)=\Gamma_0$). The high entropy cases are in the
thick shell regime ($\xi=0.1$, 0.006 for $\Gamma_0=10^3$, $10^4$,
respectively) while the low entropy case is in the thin shell regime
($\xi=3$). As a consequence, the time at which the afterglow emission
peaks does not depend on $\Gamma_0$ for the high entropy cases. This
is due to the fact that in this case the reverse shock crossing the
fireball becomes relativistic before it reaches the end of the
fireball (Sari 1997).  In these cases the forward shock light curve
peaks at $t\simeq{}t_\gamma+10$~s, roughly coincident with the time
(in the GRB rest frame) when half the energy has been released by the
central engine.

In all three cases we find that a simple power-law fit is obtained
fixing $t_0=t_\gamma$. This implies that the steep decay derived by
Tagliaferri et al. (2005) is not affected by a bad choice of the time
reference, as some authors have suggested (Tagliaferri et al. 2005,
Chincarini et al. 2005; Panaitescu et al. 2005; Zhang et
al. 2005). Tagliaferri et al. (2005) model the decay with a broken
power-law. Most of the models suggested in the literature (Panaitescu
et al. 2005; Zhang et al. 2005) invoke two separate components for the
early and late time afterglow, respectively. We therefore model the
light curve as the sum of two power-laws as well as a broken
power-law. We also explore the possibility that the early phase can be
modeled with an exponential decay rather than a power-law.

\begin{figure}
\plotone{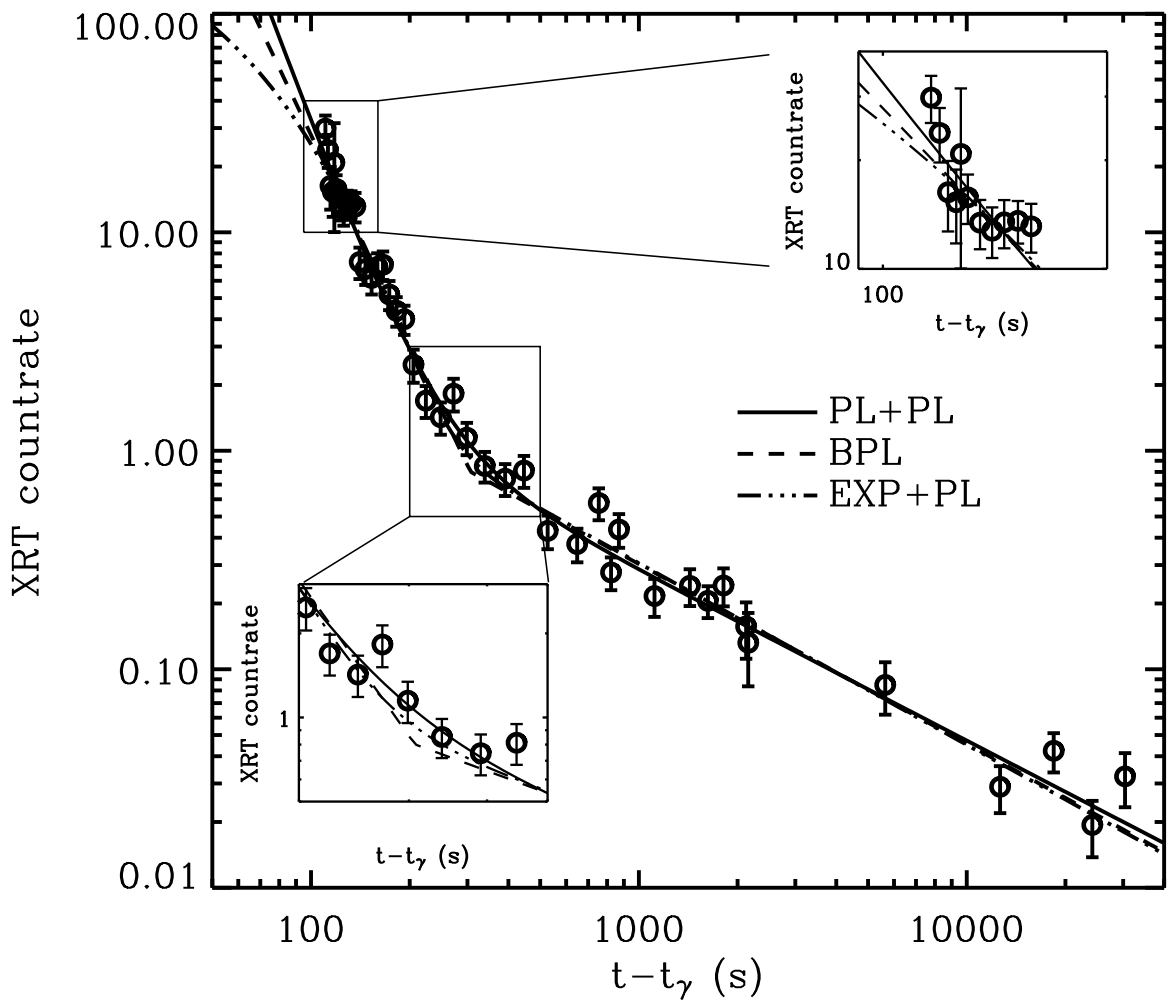} 
\figcaption{The Swift XRT light curve of GRB~050219A. The three best
fit models discussed in text(see also Table~\ref{tab:2}) are
overlaid. The two insets show expanded time intervals where the
difference between the three models is more evident: the early times
and the ankle region.
\label{fig:data}}
\end{figure}

We first consider the case in which the early steep decay of the
afterglow corresponds to the interaction of the fireball with the
external material. In this case, as discussed above and shown in
Fig.~\ref{fig:lc0219}, we can assume $t_0=t_\gamma$ and, after
marginalizing over the late afterglow slope and normalization as well
as the early afterglow normalization, all the fits depend only on one
parameter, i.e., the early time slope $\delta_1^X$. We show in
Table~\ref{tab:2} the result of the fits. The best model is the sum of
two power-laws, which is valid for any physical model in which the
steep and shallow decaying portions originate in different places
(e.g., forward and reverse shock) or by different processes (e.g.,
synchrotron and inverse Compton). However, the steepness of the decay
is not consistent with any of those physical explanations. The broken
power-law model yields a shallower decay but has a significantly worse
$\chi^2$. It also has little physical support since any sharp feature
in the intrinsic emission would be smeared out by the fireball
curvature and result in an observed profile with a smooth transition
from the steep to the shallow decay slopes (Lazzati et
al. 2002). Figure~\ref{fig:data} shows the three best fit models with
the regions where the models differ zoomed in.  We do not report here
the smoothly broken power-law fits we performed since the additional
free parameter involved makes all constraints looser. In addition, the
F-test shows that the additional parameter is not required if compared
to the sum of two power-laws. We therefore conclude that the X-ray
data of GRB~050219A do not support any interpretation that explains
the steep decay portion of the afterglow as due to the interaction
with the fireball with the external medium, irrespective of the
stratification of the ISM (Panaitescu et al. 2005). We finally note
that a model with an exponential decay plus a power-law provides
almost as good a fit as the sum of two power-laws with the same number
of free parameters. We will consider this model in more detail below.

Next we consider a model in which the fireball is fragmented into
bullets (Heinz \& Begelman 1999; see also Dar \& De Rujula 2004 and
Yamazaki et al. 2004 for analogous geometries). In this case both the
prompt and the afterglow emission originate from the interaction of
the bullets with the ISM. Since the bullets are causally connected,
they begin expanding sideways almost immediately after the
deceleration and the initial afterglow is as steep as that of a normal
fireball after the jet break ($F(t)\propto{}t^{-p_e}$). Since every
bullet hits a fresh portion of the ISM, it is possible that the
dominant bullet has $t_0>t_\gamma$ and no constraint on $t_0$ can be
derived from our simulations. The best guess on $t_0$ is given by the
beginning of the last spike in the prompt emission light curve. This
is due to the fact that, if all spikes decay in time with the same
profile, the emission from the last visible spike dominates the
combined emission from all other spikes at later times. In the case of
GRB~050219A (see Fig. 2 of Tagliaferri et al. 2005) the light curve
shows a single broad peak, but it is not possible to rule out that the
smooth profile is due to the overlap of many narrower pulses. The best
constraint we can obtain is $t_0-t_\gamma\lesssim35$~s. For this
reason we allow for a variable $t_0-t_\gamma$ in the PL+PL
fits. Fig.~\ref{fig:at0} shows the confidence regions for the two
parameters. They are, as expected, highly correlated. Not
surprisingly, given the discussion above, the best fit $t_0$ is
located at a time that precedes the trigger by about 7~s. The
$1-\sigma$ contour includes a decay slope as shallow as
$\delta_1^X=2.6$ which allows freedom of interpretation. Such shallow
decays would require $t_0-t_\gamma\gtrsim40$~s, somewhat larger than
our estimate above. This is not impossible, since the early X-ray
afterglow could originate from a weak low peak-energy spike not
detected in the Swift BAT. A similar suggestion comes from the
mismatching fluxes in Fig. 3 of Tagliaferri et al. (2005). The slope
of the electron distribution can be derived from the spectral fits of
Tagliaferri et al. (2005), yielding $p_e=2.2\pm0.4$. A comparison with
Fig.~\ref{fig:at0} shows that this model is marginally consistent with
the data.

Finally we consider models in which the steep decaying radiation is
associated with the same mechanism that produces the prompt emission
(Kumar \& Panaitescu 2000; Zhang et al. 2005; Fan \& Wei 2005; Nousek
et al. 2005). In this case the early X-ray afterglow is nothing more
than the radiation produced during the prompt phase by portions of the
fireball that move at angles $\theta\gg1/\Gamma$ from the line of
sight. Again in this case the two slope sections must arise from
different locations and so the BPL fits are not relevant.  The large
angle emission from the fireball has a decay law
\begin{equation}
\delta^X=-2-\alpha^X=-3.1\pm0.2
\end{equation}
(Kumar \& Panaitescu 2000), where $\alpha^X$ is the X-ray spectral
slope and the time is measured from the beginning of the associated
prompt emission pulse. If the time is measured from the beginning of
the burst, a steeper slope can be measured (e.g. Fan \& Wei 2005),
analogously to what discussed above for the external shock. The
observational constraint is obtained by Tagliaferri et al. (2005)
fitting the X-ray spectrum. This observational constraint appears as a
gray shaded area in Fig.~\ref{fig:at0}, showing the consistency of the
data with this interpretation. We favor this interpretation over the
bullets because the predicted slope is steeper and therefore it is not
necessary to stretch the $t_0-t_\gamma$ value to the margin of the
allowed region.

The latter interpretation carries important consequences. The fact
that the steep decay is observed for $\gtrsim250$~s and that
$t_0\lesssim50$~s imply
\begin{equation}
(1+z)\frac{2R_\gamma}{c}(1-\cos\theta_j)\gtrsim200\;\;{\rm s,}
\label{eq:rg}
\end{equation}
where $\theta_j$ is the jet opening angle. This results in a
constraint on the radius $R_\gamma$ at which the prompt emission
photons are produced:
\begin{equation}
R_\gamma\gtrsim\frac{1.2\times10^{13}}{(1+z)(1-\cos\theta_j)} \qquad
{\rm{cm}}
\end{equation}
which, for a beaming angle of 10 degrees (larger than average;
Ghirlanda et al. 2004), yields a prompt emission radius
$R_\gamma\gtrsim4\times10^{14}$~cm. This is somewhat larger than that
inferred from theory (Rees \& \mes\ 1994; Lazzati et al. 1999).
Another important conclusion is that the high energy power-law in the
prompt emission spectrum must extend to very high frequencies, since
we still observe a power-law behavior after de-beaming.

A better case for constraining the radius at which the prompt
radiation is released is provided by GRB~050315, for which both
redshift and a possible jet break are identified (Vaughan et
al. 2005). This burst does not have the same high quality early time
data as GRB~050219A, but shares the same behavior. Combining the
redshift and energetics information we obtain a jet opening angle
$\theta_j=6.3\,n^{1/8}$ degrees (using eq.~1 of Ghirlanda et
al. 2004). Since the steep decay phase lasts at least $100$~s in the
comoving frame after the end of the prompt phase, we obtain from
eq.~\ref{eq:rg}:
\begin{equation}
R_\gamma>2.5\times10^{14} \quad {\rm{cm}},
\end{equation}
where the $\gamma$-ray efficiency has been set to 0.2 and the density
of the ISM is $n=1$~cm$^{-3}$.

We finally comment on the possible exponential fit to the initial
decay of the X-ray afterglow. In principle this may be interpreted as
late time emission from the prompt phase due to a cooling population
of electrons. The radiation is produced at small angle
($\theta\le1/\Gamma$) in contrast to the large angle radiation
discussed above. The shock accelerated electrons cool due to radiative
and adiabatic losses. If adiabatic losses dominate and the shell
volume scales as the inverse of the square of the shell radius, the
Lorentz factor of the electrons evolves as
$\gamma_e\propto{R}^{-2/3}$, where a relativistic equation of state
with adiabatic index 4/3 has been assumed. If the magnetic field
remains in equipartition, the flux at frequencies above the synchrotron
frequency will decay as
\begin{equation}
F=F_\gamma\exp\left[-\frac{\nu}{\nu_\gamma}\left(\frac{R}{R_\gamma}
\right)^{7/3}\right],
\end{equation}
where the subscript ``$\gamma$'' indicates quantities measured at the
radius at which the prompt radiation is produced. The above equation
gives a super-exponential decay with an e-folding time
$\tau=R_\gamma/(\Gamma^2\,c)$. As a consequence, the measured
e-folding time of $\sim40$~s results in a radius
$R_\gamma=10^{16}\,(\Gamma/100)^2$~cm, comparable to the radius of a
typical external shock. Such a large radius is, however, inconsistent
with the assumption that the electrons are cooling adiabatically,
since the reverse shock would re-energize them and destroy the
exponential decay of the light curve. In addition, if we attempt to
fit the light curve with a super-exponential decay
[$F(t)\propto\exp(t^{-A})$], the fit worsens significantly for $A>1$.

\section{Discussion}

\begin{figure}
\plotone{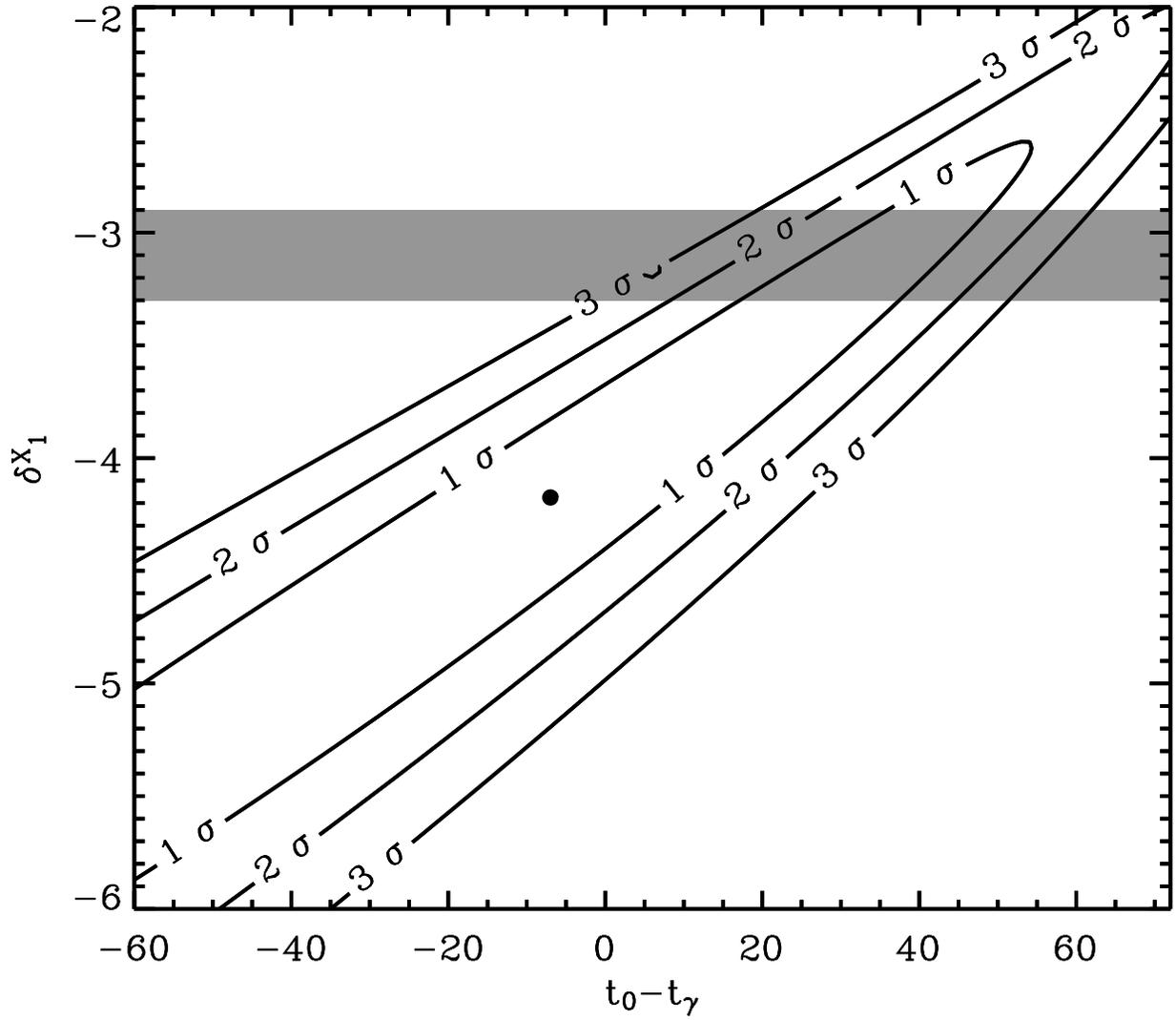} 
\figcaption{Confidence regions for the early afterglow power-law slope
($\alpha_1$) and the self-similarity start time ($t_0$) of
GRB~050219A.
\label{fig:at0}}
\end{figure}

The observation of steep decays in the early X-ray afterglows of GRBs
is one of the highlights of the Swift mission (Tagliaferri et
al. 2005; Chincarini et al. 2005). Several possible explanations for
these decays have been suggested. These include external shocks
propagating into heavily stratified media (Panaitescu et al. 2005),
reverse shock emission (Kobayashi et al. 2005), expanding bullets
(Heinz \& Begelman 1999) and large-angle emission from the prompt
phase (Kumar \& Panaitescu 2000). A better understanding of the
phenomenon has been hampered by the difficulty in pinning down the
decay slope of the early phase since the early slope is highly
correlated with the assumed reference time $t_0$. Changing $t_0$,
which potentially lies anywhere within the prompt emission episode,
can modify the early decay law substantially (see Fig.~\ref{fig:at0}).
A late time $t_0$ was used in the past (e.g. Piro et al. 2005) to
explain late X-ray bumps as the beginning of the afterglow. We now
know that these bumps are most likely late episodes of central engine
activity (Burrows et al. 2005; Falcone et al. 2005).

The above-mentioned models can be divided into two main classes: those
that explain the early afterglow as radiation from the interaction of
the fireball with the surrounding medium (external or reverse shocks)
and those that associate the early afterglow with the tail of the
prompt phase. We first considered the early afterglow as a
manifestation of the fireball/ISM interaction. We simulated early
external shock emission from a thick fireball. In most cases,
measuring times from the beginning of the prompt phase
($t_0=t_\gamma$) is a very good approximation and does not lead to an
overestimate of the early-time slope. Only for light curves in which
the emission is concentrated at the end of $T_{\rm{GRB}}$ do we find
evidence of deviations from a pure power-law decay if we assume
$t_0=t_\gamma$. We can therefore eliminate one of the main unknowns in
the modeling of early afterglow data, at least as far as ISM models
are concerned. Fixing $t_0=t_\gamma$, we find that the early decay
slopes are too steep to be explained within the fireball model. This
is explicitly discussed for the case of GRB~050219A, but analogous
conclusions hold for most of the bursts discussed by Tagliaferri et
al. (2005) and Chincarini et al. (2005).

There are two situations, however, in which the reference time $t_0$
can be larger than $t_\gamma$. If the fireball is fragmented into
bullets, then $t_0$ is given by the ejection time of the dominant
bullet, while if the early X-ray afterglow radiation is associated
with the tail of the prompt phase, $t_0$ is the beginning of the last
bright episode in the prompt light curve.  Focusing again on the light
curve of GRB~050219A, we show that the data are consistent with both
models. Bullets, however, are only marginally consistent with the
data, since the predicted decay slope is shallower than that of
large-angle radiation from the prompt phase (see also Zhang et
al. 2005). We discuss the constraints provided by this interpretation
on the radius at which the prompt radiation is produced. GRB~050219A
does not have an estimated beam opening angle and as a consequence it
is not possible to place robust limits on the emission radius. We
therefore discuss the limits obtained from GRB~050315, for which both
a redshift and a break time have been measured (Vaughan et
al. 2005). For this burst we find a constraint
$R_\gamma\ge2.5\times10^{14}$~cm, a somewhat larger radius than
expected (Rees \& \mes\ 1994; Lazzati et al. 1999). Within the
internal shock scenario, such a large radius would imply either a
large Lorentz factor ($\Gamma_0>3000$ for shells separated by 1~ms) or
a long interval between the ejection of two shells (1~s for
$\Gamma_0=100$). Future observations increasing the sample of GRBs
with steep decay, redshift measurements, and measurements of jet breaks
will help to pin down this important radius more precisely. We also
showed that a formally acceptable fit can be obtained with an
exponential decay instead of a steep power-law. The best-fit e-folding
time is, however, too long to be associated with any cooling time in
the fireball and therefore a physical interpretation is not
straightforward.

\bigskip

We thank G. Tagliaferri, A. Moretti, S. Campana and the Swift team for
providing us with the X-ray light curves of GRB~050219A and GRB~050315
and for useful discussions. This work was supported by NSF grant
AST-0307502 and NASA Astrophysical Theory Grant NAG5-12035.

\clearpage

\begin{deluxetable}{ccc} 
\tablecolumns{3} 
\tablewidth{0pc} 
\tablecaption{Results from the fits. \label{tab:1}} 
\tablehead{$\alpha$ & $\frac{t_0-t_\gamma}{T_{\rm{GRB}}}$ optical & 
$\frac{t_0-t_\gamma}{T_{\rm{GRB}}}$ X-ray}
\startdata 
-1   & -0.5    & -0.25  \\
0    & -1.0    & -0.6   \\
1    & -0.85   & -0.45  \\
10   & 0.3     & 0.55  \\
100  & 0.9     & 0.9   \\
\enddata 
\end{deluxetable}

\begin{deluxetable}{cccc} 
\tablecolumns{4} 
\tablewidth{0pc} 
\tablecaption{Results of the fits to the early decay of the X-ray afterglow
of GRB~050219A with the three models discussed in the
text. $t_0=t_\gamma$ is assumed.\label{tab:2}} 
\tablehead{Model & $\delta_1^X$ ($90\%$ c.l.) & $\chi^2/$d.o.f. & Probability}
\startdata 
BPL   & $-3.2\pm0.3$               & 56.1/39         & 0.04 \\
PL+PL & $-4.0\pm0.4$               & 49.9/39         & 0.12 \\
Exp+PL& $39\pm5$\tablenotemark{a}   & 54.4/39         & 0.05 \\
\enddata 
\tablenotetext{a}{For the exponential plus power-law case, the
e-folding time is reported instead of the early time slope.}
\end{deluxetable}

\clearpage

\end{document}